\documentclass[runningheads,a4paper]{llncs}

\usepackage{amssymb}
\usepackage{graphicx}
\usepackage{multirow}
\usepackage{rotating}
\usepackage{url}
\newcommand{\keywords}[1]{\par\addvspace\baselineskip
\noindent\keywordname\enspace\ignorespaces#1}
\urldef{\mailsa}\path|{aybuke.ozturk, stephane.lallich, jerome.darmont}
@univ-lyon2.fr|
\urldef{\mailsb}\path|{louis.eyango, yona.waksman}
@mom.fr|

\begin{document}

\mainmatter  % start of an individual contribution

\title{Warehousing Complex Archaeological Objects}

\titlerunning{Warehousing Complex Archaeological Objects}

\author{Ayb\"uke \"Ozt\"urk$^{*, **}$
\and Louis Eyango$^{**}$
\and Sylvie Yona Waksman$^{**}$ 
\and \\ St\'ephane Lallich$^{*}$
\and J\'er\^{o}me Darmont$^{*}$}
\authorrunning{Warehousing Complex Archaeological Objects}

\institute{Universit\'e de Lyon, France\\
$^{*}$Laboratoire ERIC (5 avenue Pierre Mend\`es France, 69676 Bron Cedex), \\
$^{**}$Laboratoire Arch\'eom\'etrie et Arch\'eologie (7 rue Raulin, 69365 Lyon Cedex 7)\\
\mailsa\\
\mailsb}


\toctitle{Warehousing Complex Archaeological Objects}
\tocauthor{Ayb\"uke \"Ozt\"urk, Louis Eyango, Yona Waksman, St\'ephane Lallich and J\'er\^{o}me Darmont}
\maketitle

\begin{abstract}
Data organization is a difficult and essential component in cultural heritage applications. Over the years, a great amount of archaeological ceramic data have been created and processed by various methods and devices. Such ceramic data are stored in databases that concur to increase the amount of available information rapidly. However, such databases typically focus on one type of ceramic descriptors, e.g., qualitative textual descriptions, petrographic or chemical analysis results, and do not interoperate. Thus, research involving archaeological ceramics cannot easily take advantage of combining all these types of information. 

In this application paper, we introduce an evolution of the Ceramom database that includes text descriptors of archaeological features, chemical analysis results, and various images, including petrographic and fabric images. To illustrate what new analyses are permitted by such a database, we source it to a data warehouse and present a sample on-line analysis processing (OLAP) scenario to gain deep understanding of ceramic context.

\keywords{\textit{Archaeology, Archaeometry, Ceramics, Complex objects, Da-tabases, Data Warehouses, OLAP}}
\end{abstract}

\section{Introduction}

Archaeology is a branch of humanities that investigates past societies. It includes the study of material culture left behind by past human populations, especially pottery, which is seen as the most common archaeological material, providing with information on many aspects including chronology, trade, and technology. The form and decoration of pottery changed over time, which makes it a potential chronological marker, and its circulation is an indication of exchanges and trade. Another interesting fact about pottery is that, once a pottery was broken, it could not be recycled, unlike iron or glass, for instance. Thence, potteries have remained to exist until today. Therefore, it is one of the most important archaeological material to help reconstruct past civilizations.

In recent years, the use of digital systems and tools in archaeological studies developed rapidly. Using web-based databases, geographical coordinates, digital mapping, and digital photography is becoming popular. Scientific developments and statistical techniques have further contributed to the analysis of archaeological materials. Nowadays, digital systems are needed by archaeologists to study a variety of archaeological information and to share them. Digital systems also give an opportunity to study different aspects of the same objects or categories of objects. 

Archaeological ceramics\footnote{In this paper, we use both ``pottery'' and ``ceramic'' to designate all the range of categories of these archaeological objects.} can be described in different ways, by archaeologists, museum curators or archaeological scientists, e.g., through chemical, mineralogical, and petrographic analyses. In addition, ceramics can be used to determine contextual relationships, which help to highlight archaeologically meaningful data from the mass of individual data. In other words, exploiting ceramic data allows to discover patterns that are only visible in larger and more distributed ceramic samples than can be collected about any single ceramic. In archaeology, core data are highly contextual. Thence, ceramics and their properties can help to obtain comprehensive knowledge about technological, cultural and geographical information. Such information may also contribute to understand the period and provenance of ceramics. 

However, information is globally very heterogeneous. Databases have different file formats, access protocols, and use various query languages. There is no standardized terminology, especially in terms of the description of ceramic materials and their properties. Moreover, databases have a strong focus by and large. For instance, in Lyon archaeometric studies carried out on ceramics \cite{Picon1993,Waskman2} led to the development of the  Ceramom database \cite{ceramom}. In Ceramom, whose development began in the late 1970's, ceramics were until recently mainly described by their chemical composition together with a text summary of archaeological information. Eventually, databases little interoperate, most being offline and the others only providing a web interface, but no API (Application Programming Interface). Thus, combining various information about archaeological objects, such as textual, numerical, and graphical documents, which would allow powerful computer analyses, is at best an intricate task as of today.

Thus, in this application paper, we introduce the new Ceramom database, which models previously little-exploited textual descriptions of ceramic samples and includes image descriptions (technical drawings, photos, etc.) as well. This new database aims to be the basis of powerful analyses, such as OLAP and data mining, which should integrate various points of view on ceramic objects, e.g., text descriptors, chemical analysis results, technical drawings, binocular images resulting from fabric analyses, and petrographic images resulting from petrographic analyses to be able to learn deep contextual information from all the bits and pieces of ceramic. To this aim, we use the new Ceramom database to source a data warehouse, which is original in its storage of data that are not only numerical. 

The remainder of this paper is organized as follows. Section~\ref{sec:CeramicDatabaseProjects} presents a selection of ceramic database projects, including Ceramom. Section~\ref{sec:TheCeramomDatabase} further details the new Ceramom database. Section~\ref{sec:ExploringArchaeologicalCeramicData} deals with remodeling this database as a multidimensional, data warehouse schema, presents a sample OLAP analysis scenario and discusses issues in ceramic data analysis. Finally, we conclude this paper and hint at future research in Section~\ref{sec:ConclusionAndPerspectives}.

\section{Ceramic Database Projects}
\label{sec:CeramicDatabaseProjects}

In recent years, several databases were created to highlight different perspectives in pottery research. These databases have different types of contents, depending on the aspects of ceramics studies they focus on. Moreover, specific formats may be implied based on different contents, e.g., numbers for chemical analyses or text and/or images for petrographic analyses.

Databases usually have a main type of contents, and may focus on specific categories of ceramics, time periods or regions. Databases can be either publicly available online or not. Additionally, interface features may also be available, such as interactive maps, interactive 2D or 3D views, or statistical tools, the latter being of particular interest in the context of our research. Table~\ref{tab:databases} lists a selection of databases that we consider representative of this diversity of contents, formats, statuses and features, with some indications on their specificity. In Table~\ref{tab:databases}, primary content is indicated by X, secondary content by x, and occasional content by (x).

\begin{table}[hbt]
\centering
\caption {Ceramic database features}
\begin{tabular}{|l|c|c|c|c|c|c|c|c|c|c|}
\hline
\multirow{2}{*}{Ceramic databases} & \multicolumn{4}{c|}{Database type} & \multicolumn{3}{c|}{Data type} & \multicolumn{3}{c|}{Features}  \\ \cline{2-11}
 & \multicolumn{1}{c|}{\begin{turn}{90}Archaeological \end{turn}} & \begin{turn}{90} Chemical \end{turn} & \multicolumn{1}{c|}{ \begin{turn}{90}Petrographic \end{turn}} & \multicolumn{1}{c|}{\begin{turn}{90}Fabric \end{turn}} & \multicolumn{1}{c|}{\begin{turn}{90}Textual \end{turn}} & \begin{turn}{90}Numerical \end{turn} & \multicolumn{1}{c|}{\begin{turn}{90}Multimedia \end{turn}} & \multicolumn{1}{c|}{\begin{turn}{90}Online\end{turn}} & \multicolumn{1}{c|}{\begin{turn}{90}Structured\end{turn}} & \begin{turn}{90}Stat tools \end{turn}   \\ \hline
LCP & X &  & x & x & x &  & x & x & x &   \\ \hline
Roman Amphorae & X &  & X & X & x &  & x & x & x &   \\ \hline
POTSHERD & X &  &  & x & x &  & x & x &  &   \\ \hline
Worcestershire Ceramics & X &  & (x) & X & x &  & x & x & x &   \\ \hline
NRFRC & (x) &  & X & X & x &  & x & (x) &  &   \\ \hline
FACEM & x &  &  & X & x &  & x & x & x &   \\ \hline
Petrodatabase & (x) &  & X & x & x &  & x & (x) & x &   \\ \hline
ICERAMM & X &  &  & x & x &  & x & x &  &   \\ \hline
PECL & X &  & (x) & (x) & x &  & x & x & x &   \\ \hline
%Beazley Archive & x &  &  &  & x &  & x & x & x & x  \\ \hline
ASCSA & X &  &  &  & x &  & x & x & x &   \\ \hline
Sgraffito in 3D & x &  &  &  & x &  & x & x &  &  \\ \hline
CeraDAT & (x) & X &  &  & x & x &  & x & x & x  \\ \hline
MURR & (x) & X &  &  & x & x &  & x &  & downloadable \\ \hline
Fribourg & (x) & X &  &  & x & x &  & x &  &   \\ \hline
Ceramom 2.0 & x & X & (x) &  & x & x &  &  & x & x  \\ \hline
Ceramom 3.0 & X & X & x & x & x & x & x & x & x & x  \\ \hline 
\end{tabular} 
\label{tab:databases}
\end{table}

The \emph{Levantine Ceramics Project} (LCP), directed by Boston University, proposes an archaeological database focusing on ceramic wares produced in the Levant, from the Neolithic to the Ottoman periods \cite{LCP}. It mainly includes archaeological data (typological, chronological, and geographical), but also provides with fabric and petrographic data. The format of LCP data is either text or image. LCP is an open, interactive internet resource. \emph{Roman Amphorae: a digital resource}, proposed by the University of Southampton, provides an online introductory resource for the study of Roman amphorae, based on a rich corpus of archaeological information together with petrographic and fabric data \cite{RomanAmphorae}. \emph{POTSHERD}  is a collection regarding pottery from the Roman period (1$^{st}$ cent. BC -- 5$^{th}$ cent. AD) in Britain and Western Europe, including distribution maps and links to complementary resources \cite{POTSHERD}. The \emph{Worcestershire On-line Ceramic Database}  was designed to make available the complete pottery fabric and form type series for Worcestershire, from the Neolithic to the early post-medieval period \cite{Worcestershire}. 

Other types of databases are more centered on either  archaeological data or petrographic and fabric data. The \emph{National Roman Fabric Reference Collection} (NRFRC) is the online version of a reference book providing detailed and standardized fabric descriptions of Roman wares found in Britain \cite{National}. \emph{FACEM} focuses on fabric data of Greek, Punic, and Roman pottery in the Southern Central Mediterranean area \cite{FACEM}. FACEM includes interactive maps and allows downloading detailed information. \emph{Petrodatabase} is a petrographic relational database proposing interactive maps \cite{Petrodatabase}. \emph{ICERAMM}  focuses on medieval and modern ceramics in western and northern France, Belgium, and Switzerland \cite{ICERAMM}. \emph{PECL} is a project of encyclopedia for ceramics of the Mediterranean and sub-Saharian region of all periods, including detailed archaeological contexts information \cite{PECL}. \emph{ASCSA.net} presents archaeological objects and contexts from the excavations of the American School of Classical Studies at Athens in the Athenian Agora and in Corinth \cite{ASCSA}. 

There are several image databases that are designed for a larger audience, using digital representations of ceramics. One of them, \emph{Sgraffito in 3D},  proposes 3D reconstructions of late medieval pottery collection from the Museum Boijmans Van Beuningen (Rotterdam, the Netherlands) \cite{Sgraffito}. 

Yet other databases focus on chemical data. The \emph{CeraDAT project}, developed by the Demokritos National Centre of Science Research (Athens, Greece), is a prototype relational database including interactive maps and focusing on the Aegean and the wider Eastern Mediterranean Region \cite{CeraDAT}. The \emph{Archaeometry Laboratory Database} of the MURR laboratory (Missouri, USA)  presents chemical analysis of ceramic artifacts from many regions, including Northern, Central and Southern America, and the Mediterranean \cite{MURR}. It also gives access to ``historical'' chemical databases, such as the Berkeley laboratory's. Archaeological information of MURR data is presented as a bibliography. The Archaeometry Research group at the University of Fribourg (Switzerland)  has established several reference groups with the chemical composition of ancient ceramics from Switzerland, Italy, France, and Germany \cite{Fribourg}. Archaeological information is also presented as a bibliography. 

Eventually, Lyon's \emph{Ceramom database} used to be mainly a chemical database \cite{Picon1993,Waskman2}, including only limited archaeological information. The new Ceramom 3.0 model we detail in Section~\ref{sec:TheCeramomDatabase} has been recentered on ceramic objects, and enriched in archaeological and multimedia contents. It covers all periods. Ceramom is not available online yet, but it will be soon at the following address \cite{ceramom}. Table~\ref{tab:databases} clearly shows that Ceramom 3.0  is one of the most comprehensive database model for ceramic data, including enriched archaeological (geographical and graphical descriptions) and archaeometric (numerical values of various analysis results) data.

\section{The Ceramom Database}
\label{sec:TheCeramomDatabase}

Data may be qualified as complex if they are  \cite{Complex}: 
\begin{itemize}
\item multiformat, i.e., represented in various formats (databases, texts, images, sounds, videos…);
\item and/or multistructure, i.e., diversely structured (relational databases, XML documents repository…);
\item and/or multisource, i.e., originating from several different sources (distributed databases, the Web…);
\item and/or multimodal, i.e., described through several channels or points of view (radiographies and audio diagnosis of a physician, data expressed in different scales or languages…);
\item and/or multiversion, i.e., changing in terms of definition or value (temporal databases, periodical surveys…).
\end{itemize}

\begin{figure}[hbt]
\begin{center}
\includegraphics[width=7cm]{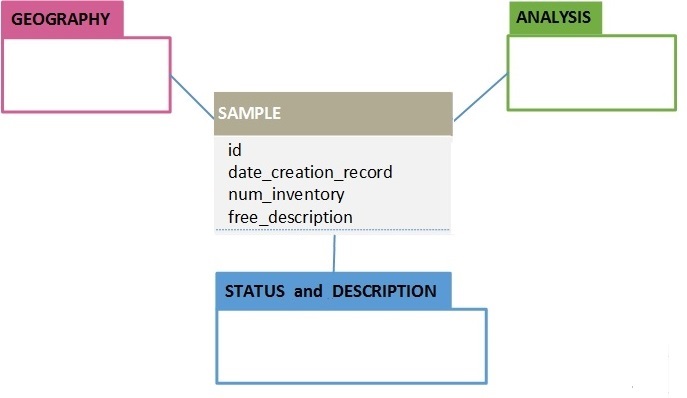}
\caption{Ceramom conceptual schema: global view}
\label{big_picture2}
\end{center}
\end{figure}

The Ceramom database was designed from the requirements of ceramic specialists for recording, using, and analysing data generated by different techniques. Thence, it stores complex data. Mineralogical and petrographical data with extensive definitions are indeed combined with rich location data. In addition to these data, graphical documents that are necessary to complement ceramic information are added in the new model, such as drawings and images of ceramic samples with variety of references. According to this model, the Ceramom database is centered on pottery samples, which are described by geographical features, several analyses, and various descriptions (Figure~\ref{big_picture2}). Each of these packages is further detailed in the following subsections. All conceptual models are depicted as UML class diagrams \cite{uml}.

\subsection{Geography}

\begin{figure}[!hbt]
\begin{center}
\includegraphics[width=12cm]{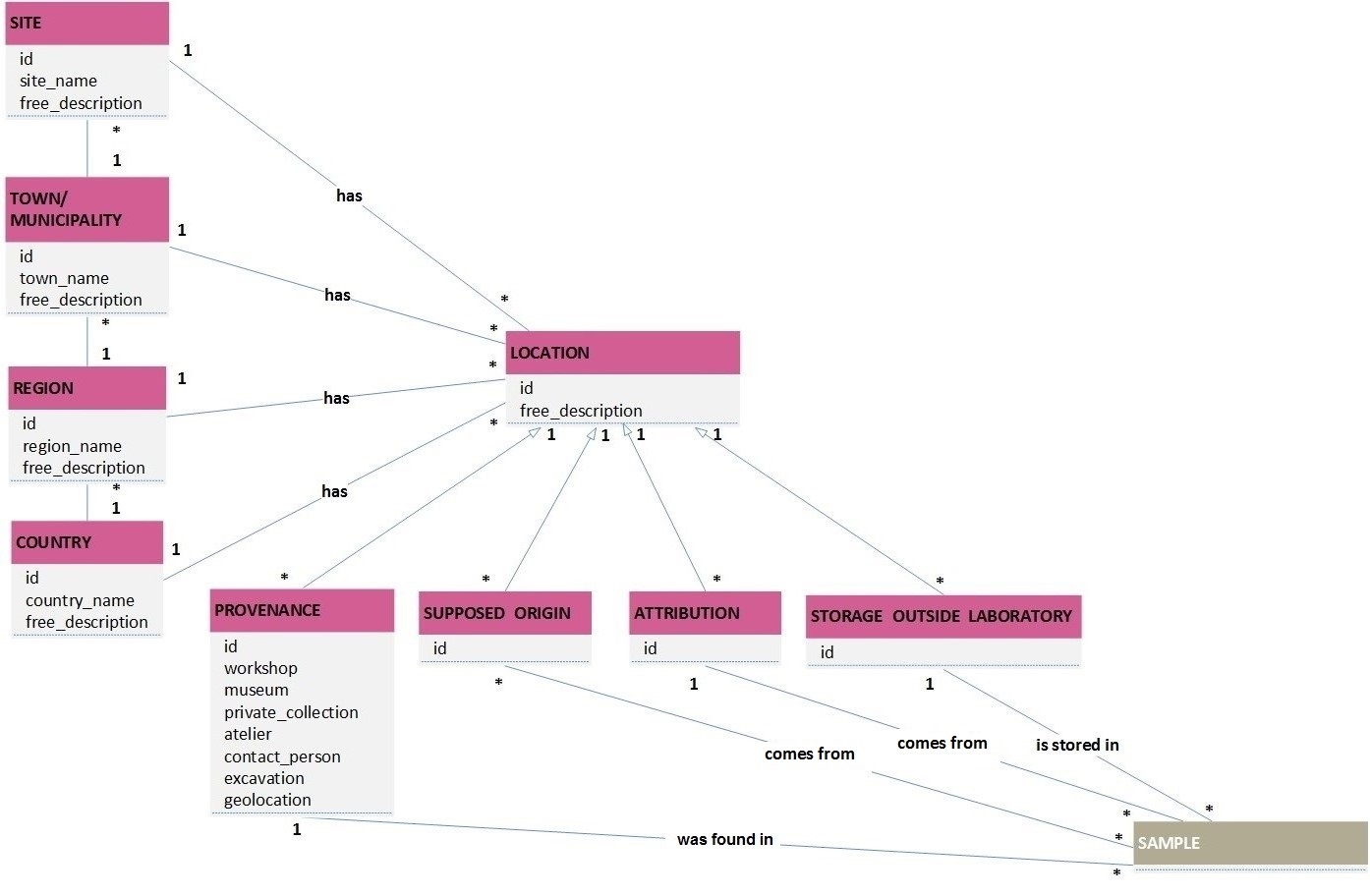}
\caption{Geography package} 
\label{Location}
\end{center}
\end{figure}

Figure~\ref{Location} displays the Geography package of the data model. The LOCATION class connects geolocation data to STORAGE OUTSIDE LABORATORY, PRO-VENANCE, SUPPOSED ORIGIN, and ATTRIBUTION classes. The STORAGE OUTSIDE LABORATORY class represents where the object is physically stored. 

The PROVENANCE class bears information regarding the location data where the object was found. The SUPPOSED ORIGIN class provides a supposed origin before analysis. The ATTRIBUTION class indicates where the object was demonstrated to come from by after analysis. Finally, the SITE, TOWN/ MUNICIPALITY, REGION, and COUNTRY classes represent a hierarchy of LOCATIONS. 

\subsection{Status and Description}

\begin{figure}[!hbt]
\centering
\includegraphics[width=12cm]{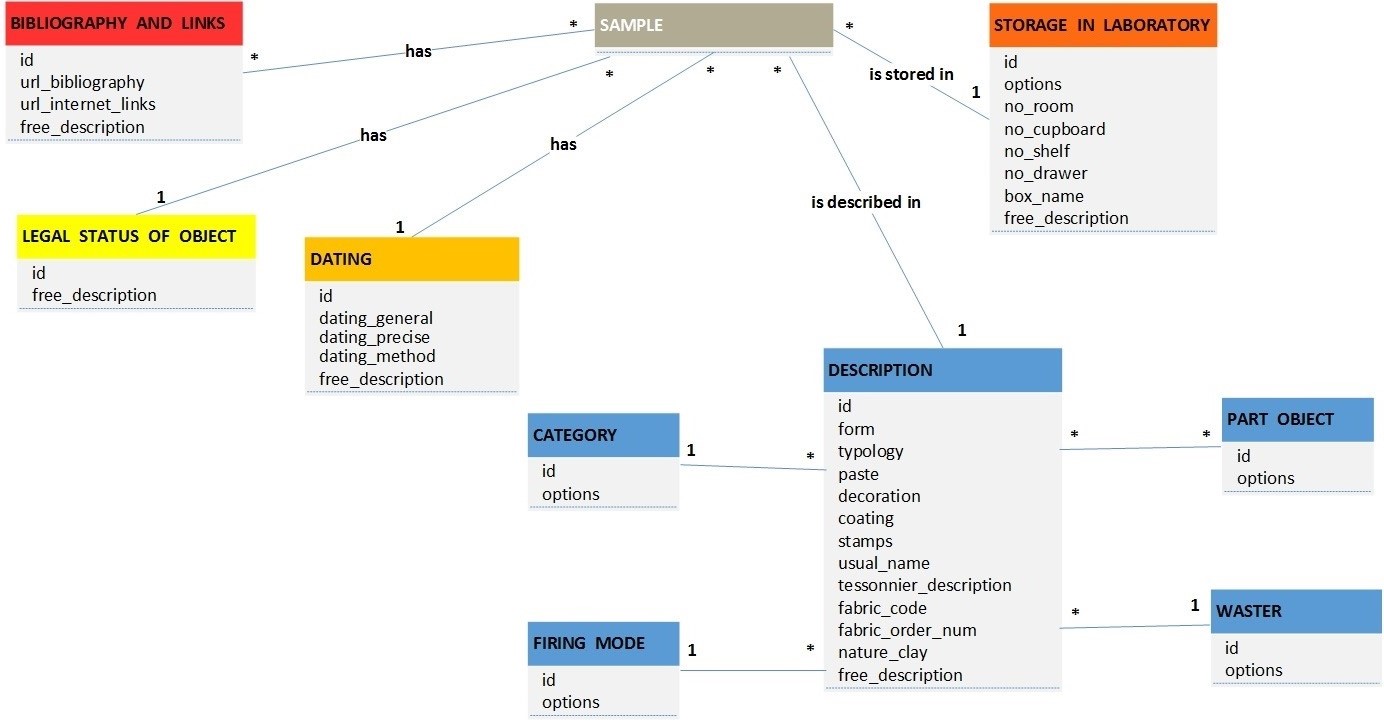}
\caption{Status and Description package}
\label{StatusAndDescription}
\end{figure}

The Status and Description package of the data model is depicted in Figure~\ref{StatusAndDescription}. The DESCRIPTION class bears textual descriptors of the object. It includes most of the information which enable us to identify the object archaeologically. The CATEGORY class helps categorize ceramics, e.g., ``COMM." (common ware), ``CARREAU" (tile). The PART OBJECT class collects data regarding parts of the form, e.g., rim, base, body, etc. The WASTER class specifies whether the object is a waster. The FIRING MODE class contains data about firing mode, which are coded as mode A, mode B, mode C, etc. The STORAGE IN LABORATORY class bears location data if the object is stored in the laboratory. The LEGAL STATUS OF OBJECT class contains data about ownership of the object. 

\subsection{Analysis}

\begin{figure}[!hbt]
\begin{center}
\includegraphics[width=9.5cm]{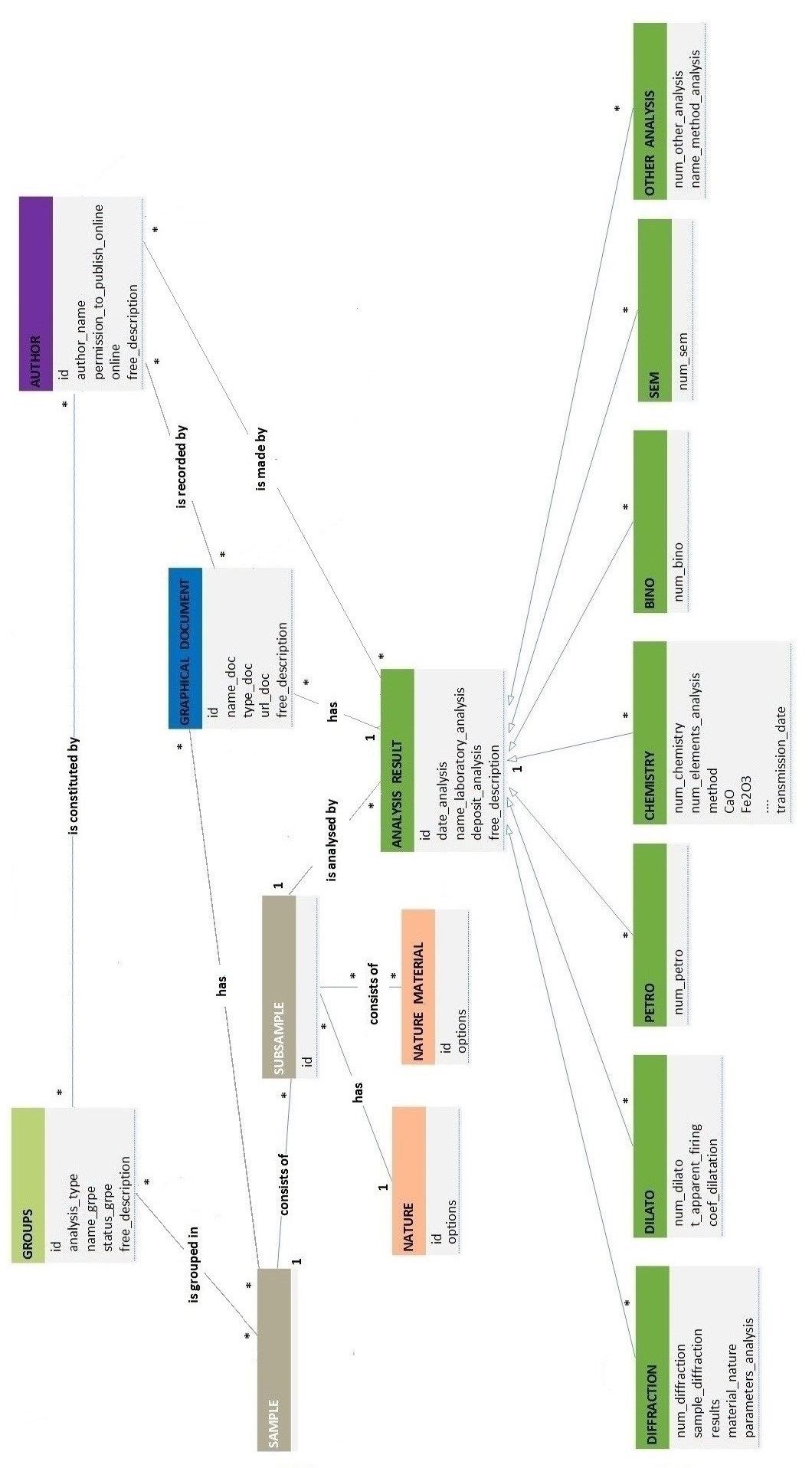}
\caption{Analysis package}
\label{Analyses}
\end{center}
\end{figure}

Figure~\ref{Analyses} depicts the Analysis package of the data model. Analysis results of a sample are represented in class ANALYSIS RESULT. Each analysis corresponds to a series of separate records, each of which contains an individual measure. This is due to the fact that samples may be analyzed by several techniques, such as chemical or petrographical analyses. A given sample may also be analyzed several times using the same technique, but with different parameters. For examples, as a larger number of chemical elements were determined since the 1970's, when a same sample was re-analyzed a given chemical element (e.g. aluminium or calcium) would be assigned several values of concentration. 

The DIFFRACTION, DILATO, PETRO, CHEMISTRY, BINO, SEM, and additional analyses classes bear data regarding diffraction, dilatometry, petrography, chemistry, binocular microscopy, scanning electron microscopy, and additional, miscellaneous analyses, respectively. 

\section{Exploring Archaeological Ceramic Data}
\label{sec:ExploringArchaeologicalCeramicData}

Data warehouses are actually databases with a specific model tailored for efficient OLAP analyses. In a data warehouse, the observed data are called facts, e.g., sales in a business context. They are characterized by measures that are usually numerical, e.g., quantities sold and amounts of money. Facts are observed with respect to different analysis axes called dimensions, e.g., sold products, store location and sale date. Thus, data warehouse schemas are called multidimensional schemas, or more casually star schemas, for facts are usually represented in the center of the model, with dimensions gravitating around. Star schemas help answer queries such as ``total sales revenue of each product in Lyon in 2014'', to go on with our business example. 

Moreover, dimensions may be organized in hierarchies, e.g., a time dimension could be subdivided into day, month, quarter, and year. Such a structure helps observing facts at different granularity levels, e.g., ``dezooming'' from one quarter of a year to said year to have a more global (aggregate) view of sales, or ``zooming'' from one month to one day in this month to have a more detailed view of sale events. These operations actually correspond to OLAP's rollup and drill-down operators, respectively.

Thus, to allow OLAP navigation in the Ceramom data, we must select facts to observe, axes of analysis (dimensions) and import data from Ceramom into the data warehouse. The result is called a cube (hypercube when the number of dimensions is greater than 3), where dimension values are coordinates that define a fact cell. There are a couple of interesting works done using OLAP analysis on archaeological data \cite{OLAP1,OLAP2}. 

\vspace{-0.5cm}
\subsection{Multidimensional Model}

\begin{figure}[hbt]
\begin{center}
\includegraphics[width=12cm]{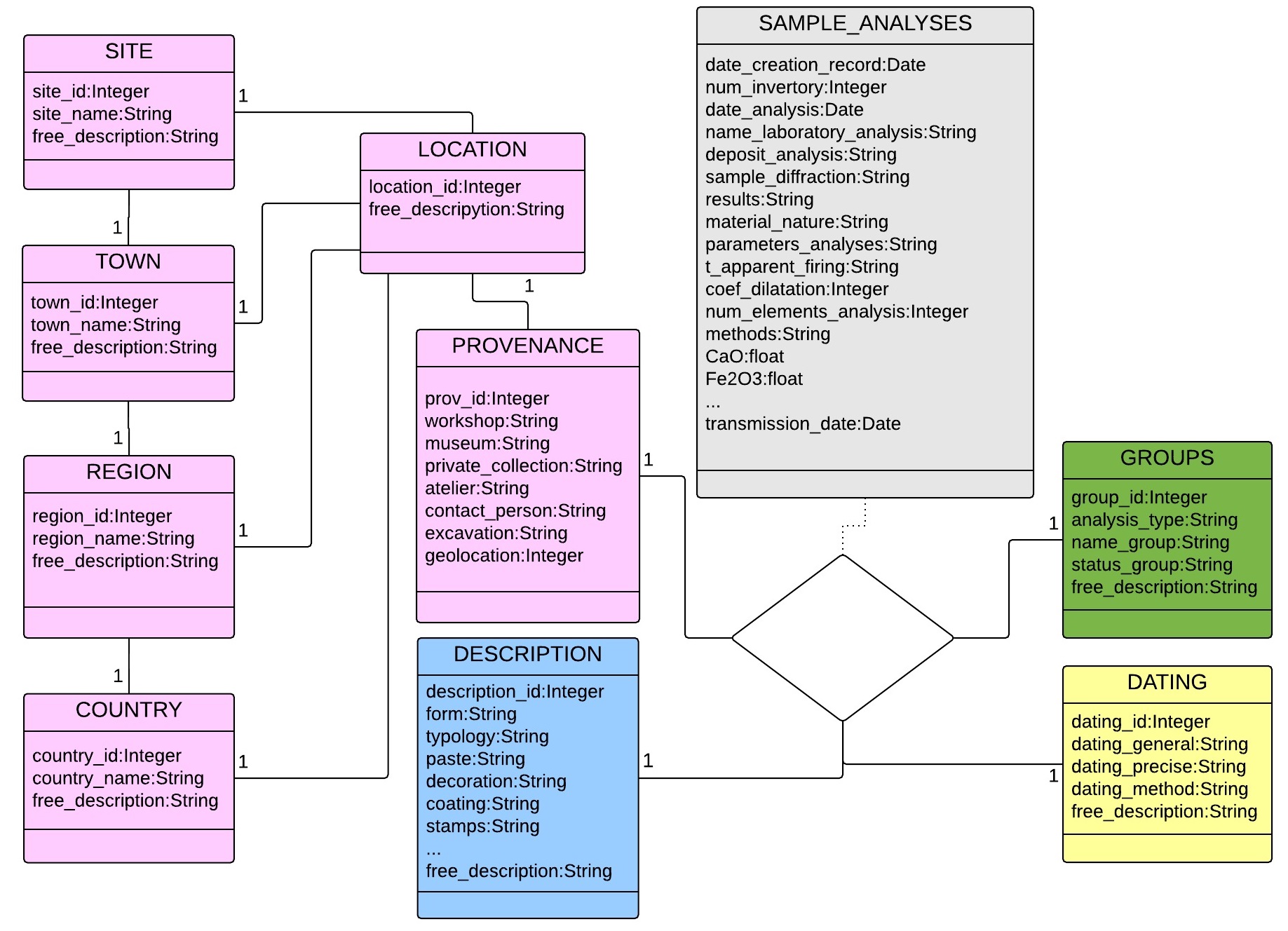}
\caption{Chemical data warehouse's multidimensional schema}
\label{UML-DW}
\end{center}
\end{figure}

In this sample scenario, we choose to observe chemical dosages with respect to ceramic sample provenance, dating, description, and groups. Our data warehouse's star schema is provided in Figure~\ref{UML-DW}, again as a UML class diagram. Facts are modeled as a quaternary association-class connected to dimension classes. To make use of numerical values for analyses, the SAMPLE class from Figure~\ref{big_picture2} is combined with the Analysis package (Figure~\ref{Analyses}) into the SAMPLE ANALYSES class in Figure~\ref{UML-DW}, which models our analysis facts. In our case, aggregates (summaries) are number of sample, average number of sample, and number of analyses, etc. Dimension classes are PROVENANCE, GROUPS, DESCRIPTION, and DATING, which are the same as in the Ceramom database (Figures~\ref{Location},~\ref{StatusAndDescription}, and Figures~\ref{Analyses}). Moreover, the LOCATION class individually connects to all classes in the SITE, TOWN/MUNICIPALITY, REGION, and COUNTRY hierarchy to still allow a connection in case of missing value at one hierarchy level (Section~\ref{issues}).

\subsection{OLAPing Archaeological Ceramic Data}

Once part of Ceramom data are multidimentionally remodeled, OLAP analyses can be done. OLAP actually helps interactively navigate the data warehouse, e.g., to discover outliers or hidden patterns. We use Pentaho Business Analytics\footnote{\url{http://www.pentaho.com}}, a suite of open source business intelligence, as our OLAP engine. Pentaho features a user console that is a web-based design environment. The console helps to visualize and navigate hypercubes, which are created from the data warehouse with the help of the schema-workbench tool. 

As an example of OLAP analysis, let us examine the contents of a specific chemical group coming from the GROUP class, and compare them to the initial typological classification from the DESCRIPTION class. Samples within a given chemical group belong to the same pottery production, i.e., they share the same origin. They usually come from several excavations (PROVENANCE class), and their circulation and corresponding fluxes provide insight into past contacts between populations and trade networks. When different workshops manufactured similar wares, classified under the same typology, chemical analysis ``sorts out'' the different productions and enables archaeologists and historians to better understand economic trends and cultural influences.

In a first analysis, successive rollups  help aggregate PROVENANCE data at the country level to achieve a coarser view of data. We take interest in ceramics of the Byzantine period (Medieval period in the DATING class) called ``Zeuxippus Ware'' and we select the DATING, DESCRIPTION (typology), and PROVENANCE (country) dimensions and count ``Zeuxippus Ware" occurrences (OLAP slice and dice operators, respectively). ``Zeuxippus Ware'' corresponds to a typological class that has 163 occurrences in the database. ``Zeuxippus Ware'' was found all over the Mediterranean and beyond, but was also largely imitated. 

In a second analysis, we slice on PROVENANCE, GROUPS, and DESCRIPTION, and dice on ``Zeuxippus Ware stricto sensu''. A research program enabled to define several distinct chemical groups, including one corresponding to the ``Zeuxippus Ware stricto sensu'' (87 samples), which is the ``prototype'' of this ware. We identify the features of each production, including information on its geographic distribution, related to trade networks \cite{Waksman}. 

\begin{figure}[hbt]
\begin{center}
\includegraphics[width=7cm]{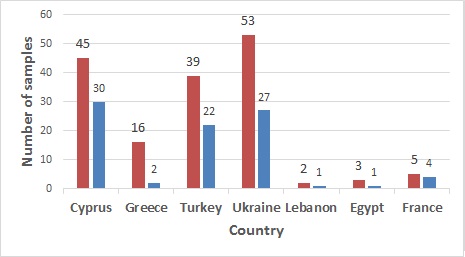}
\caption{Distribution of samples by country, for descriptions including the term ``Zeuxippus'' (in red, ``typological classification''), and in chemical group ``Zeuxippus Ware stricto sensu'' (in blue, ``chemical classification'').}
\label{fig6}
\end{center}
\end{figure}

Figure~\ref{fig6} compares data with the number of samples whose description includes the term ``Zeuxippus'', i.e., including both ``Zeuxippus Ware stricto sensu'' and wares imitating it or related to it typologically. In all countries, examples of both prototype and imitations were found. It is nonetheless noticeable that a larger proportion of imitations come from Greece, a new insight that may be significant. In histogram, ``chemical classification''(in blue) is based on the actual diffusion of ceramic products, it is related to economic factors. It confirms the large distribution of this ware in countries of the Mediterranean and Black sea areas. Although the bias introduced by the initial sampling needs to be taken into account, the number of samples from each country gives an idea of the abundance of this ware, e.g., only very few examples were found in France. ``typological classification" (in red) refers to the diffusion of models and fashions, and is thus more related to cultural factors. This example somehow simulates the comparison of data obtained on the same categories of objects from two databases, focusing each on another aspect of these wares. It shows the discrepancies, but also the added value that may be obtained when connecting information. It also shows how OLAP analysis may contribute to the understanding of economic and cultural relationships at the Byzantine period, thanks to its ability to bring ceramic information back into a wider context.

\subsection{Issues in Archaeological Ceramic Data Analysis}
\label{issues}

We have been confronted to a couple of major challenges before OLAP analyses and when performing OLAP analyses onto Ceramom data. First, we encountered a classical problem in databases, i.e., missing values. In our example scenario, there is location information for provenance studies, but in practice, some information, i.e., site, town, region or country might be missing in the database. For example, some samples relate to Sudak, Ukraine and Acre, Israel, with no archaeological site reference. This is why we complement the geographical hierarchy with direct associations from provenance to site, town, region, and country. Some of this missing information (town to region to country relationships) shall be found in external sources, though. 

Moreover, data about archaeological ceramics mostly consists of textual and numerical data. On one hand, textual data include information about the characteristics of ceramics, such as form, chronological information, etc. On the other hand, numerical data include information produced by various analyses performed on the ceramic material. Both can be warehoused. However, while classical OLAP provides a good tool for analyzing numerical data (through aggregation functions such as sum, average, minimum, maximum, etc.), it is not very convenient for textual data, whose individual values can only be counted. Thus, in order to gain in-depth knowledge about ceramics, there is a crucial need to better take textual data into account in OLAP.

\section{Conclusion and Perspectives}
\label{sec:ConclusionAndPerspectives}

Designing comprehensive archaeological databases/tools is a challenge because of many reasons. Various dimensions should be integrated from distant databases that describe the same categories of objects in a complementary way. Thus, different point of views and parameters should be combined coming from different, heterogeneous databases. 

In this paper, we first survey representative ceramic databases and show there is no single comprehensive resource for studying ceramic materials. Then, we propose a new conceptual data model for archaeological ceramics. We believe this model could be a good starting point to help ceramic databases interoperate in the future. Moreover, we illustrate how ceramic data can source a data warehouse to perform OLAP. Such analyses help navigate and observe data from different perspectives, thus enabling archaeological researchers with better insights on their data. The main benefit of the proposed approach is to identify hidden patterns and possibly unexpected values, especially only visible in larger
and more distributed ceramic samples, in order to contextualize information and help building up knowledge of past societies \cite{Kansa1,Kansa2}.

Our first perspective is to include textual and graphical descriptions of ceramic samples in the warehouse to do OLAP analyses. Some research already address non-numerical data integration into data warehouses, e.g., by performing some preprocessing on text before storage into the data warehouse. Actually, integration of such data might not be in textual format \cite{Pujolle2011,irma06,Aknouche13}. 

Moreover, data mining could also be combined to OLAP to complement data navigation with automatic pattern or structure discovery. For example, clustering techniques could help enhance current statistical tools for categorizing ceramics based on textual, numerical, and graphical information. Each context allows to create a clusterer. The challenge will be to manage the collaboration between the different clusterers either for producing a consensus clustering or to explain and annotate a given clustering or to identify strong forms. 

In the longer run, we finally aim to build smart links between ceramic databases in order to achieve interoperability between Ceramom and partner databases, and allow cross-analyses to run over a database federation.  

\section*{Acknowledgements}
\label{sec:Acknowledgements}

This project is supported by the Rh\^{o}ne Alpes Region's ARC 5: ``Cultures, Sciences, Soci\'et\'es et M\'ediations'' through a PhD grant to A. \"Ozt\"urk, who gratefully acknowledges this support. We also sincerely thank the Archaeological Ceramics team of the "Arch\'eom\'etrie et Arch\'eologie" Laboratory (CNRS UMR 5138), and especially C. Brun, for its contribution to the design of the Ceramom model.

\end{document}